\documentclass[proof]{pasj01}

\usepackage{color}

\newcounter{author}
\def\authorcount#1#2{\refstepcounter{author}\label{#1}
                     \altaffiltext{\ref{#1}}{#2}}

\begin{document}
\SetRunningHead{K. Namekata et al.}{WZ Sge-type Dwarf Nova: ASASSN-15po}

\Received{201X/XX/XX}
\Accepted{201X/XX/XX}
         
\title{Superoutburst of WZ Sge-type Dwarf Nova Below the Period Minimum: ASASSN-15po}
\author{Kosuke \textsc{Namekata}\altaffilmark{\ref{affil:KU}*},
Keisuke \textsc{Isogai}\altaffilmark{\ref{affil:KU}},
Taichi \textsc{Kato}\altaffilmark{\ref{affil:KU}},
Colin~\textsc{Littlefield}\altaffilmark{\ref{affil:LCO}},
Katsura~\textsc{Matsumoto}\altaffilmark{\ref{affil:OKU}},
Naoto~\textsc{Kojiguchi}\altaffilmark{\ref{affil:OKU}},
Yuki~\textsc{Sugiura}\altaffilmark{\ref{affil:OKU}},
Yusuke~\textsc{Uto}\altaffilmark{\ref{affil:OKU}},
Daiki~\textsc{Fukushima}\altaffilmark{\ref{affil:OKU}},
Taiki~\textsc{Tatsumi}\altaffilmark{\ref{affil:OKU}},
Eiji~\textsc{Yamada}\altaffilmark{\ref{affil:OKU}},
Taku~\textsc{Kamibetsunawa}\altaffilmark{\ref{affil:OKU}},
Enrique~de~\textsc{Miguel}\altaffilmark{\ref{affil:dem1}}$^,$\altaffilmark{\ref{affil:dem2}},
William~L.~\textsc{Stein}\altaffilmark{\ref{affil:Stein}},
Richard~\textsc{Sabo}\altaffilmark{\ref{affil:Sabo}},
Maksim~V.~\textsc{Andreev}\altaffilmark{\ref{affil:Terskol}}$^,$\altaffilmark{\ref{affil:ICUkraine}},
Etienne~\textsc{Morelle}\altaffilmark{\ref{affil:Morelle}},
E.~P.~\textsc{Pavlenko}\altaffilmark{\ref{affil:CRI}},
Julia~V.~\textsc{Babina}\altaffilmark{\ref{affil:CRI}},
Alex~V.~\textsc{Baklanov}\altaffilmark{\ref{affil:CRI}},
Kirill~A.~\textsc{Antonyuk}\altaffilmark{\ref{affil:CRI}},
Okasana~I.~\textsc{Antonyuk}\altaffilmark{\ref{affil:CRI}},
Aleksei~A.~\textsc{Sosnovskij}\altaffilmark{\ref{affil:CRI}},
Sergey~Yu.~\textsc{Shugarov}\altaffilmark{\ref{affil:Sternberg}}$^,$\altaffilmark{\ref{affil:Slovak}},
Polina~Yu.~\textsc{Golysheva}\altaffilmark{\ref{affil:Sternberg}},
Natalia~G.~\textsc{Gladilina}\altaffilmark{\ref{affil:Sternberg2}},
Ian~\textsc{Miller}\altaffilmark{\ref{affil:Miller}},
Vitaly~V.~\textsc{Neustroev}\altaffilmark{\ref{affil:Neustroev}}$^,$\altaffilmark{\ref{affil:Neustroev2}},
Vahram~\textsc{Chavushyan}\altaffilmark{\ref{affil:Neustroev3}},
Jos\'e~R.~\textsc{Vald\'es.}\altaffilmark{\ref{affil:Neustroev3}},
George~\textsc{Sjoberg}\altaffilmark{\ref{affil:George1}}$^,$\altaffilmark{\ref{affil:George2}},
Yutaka~\textsc{Maeda}\altaffilmark{\ref{affil:Mdy}},
Hiroshi~\textsc{Itoh}\altaffilmark{\ref{affil:Ioh}},
Gianluca~\textsc{Masi}\altaffilmark{\ref{affil:Masi}}, 
Ra\'ul~\textsc{Michel}\altaffilmark{\ref{affil:UNAM}},
Pavol~A.~\textsc{Dubovsky}\altaffilmark{\ref{affil:Dubovsky}},
Seiichiro~\textsc{Kiyota}\altaffilmark{\ref{affil:Kiyota}},
Tam\'as \textsc{Tordai}\altaffilmark{\ref{affil:trd}},
Arto~\textsc{Oksanen}\altaffilmark{\ref{affil:Nyrola}},
Javier~\textsc{Ruiz}\altaffilmark{\ref{affil:Ruiz1}}$^,$\altaffilmark{\ref{affil:Ruiz2}}$^,$\altaffilmark{\ref{affil:Ruiz3}},
Daisaku \textsc{Nogami}\altaffilmark{\ref{affil:KU}}
}

\authorcount{affil:KU}{Department of Astronomy, Kyoto University, Kitashirakawa-Oiwake-cho, Sakyo-ku, Kyoto 606-8502}

\altaffiltext{*}{namekata@kusastro.kyoto-u.ac.jp}

\authorcount{affil:LCO}{
     Department of Physics, University of Notre Dame, Notre Dame,
     Indiana 46556, USA}

\authorcount{affil:OKU}{
     Osaka Kyoiku University, 4-698-1 Asahigaoka, Osaka 582-8582}

\authorcount{affil:dem1}{Departamento de F\'{i}sica Aplicada, Facultad de Ciencias Experimentales, 
Universidad de Huelva, 21071 Huelva, Spain   }

\authorcount{affil:dem2}{Center for Backyard Astrophysics, Observatorio del CIECEM, Parque Dunar, 
Matalasca\~{n}as, 21760 Almonte, Huelva, Spain}

\authorcount{affil:Stein}{
6025 Calle Paraiso, Las Cruces, New Mexico 88012, USA}

\authorcount{affil:Sabo}{
     2336 Trailcrest Dr., Bozeman, Montana 59718, USA}

\authorcount{affil:Terskol}{
     Terskol Branch of Institute of Astronomy, Russian Academy of Sciences,
     361605, Peak Terskol, Kabardino-Balkaria Republic, Russia}

\authorcount{affil:ICUkraine}{
     International Center for Astronomical, Medical and Ecological Research
     of NASU, Ukraine 27 Akademika Zabolotnoho Str. 03680 Kyiv,
     Ukraine}

\authorcount{affil:Morelle}{
     9 rue Vasco de GAMA, 59553 Lauwin Planque, France}

\authorcount{affil:CRI}{Crimean Astrophysical Observatory, 298409 Nauchny, Crimea}

\authorcount{affil:Sternberg}{
     Sternberg Astronomical Institute, Lomonosov Moscow University, 
     Universitetsky Ave., 13, Moscow 119992, Russia}

\authorcount{affil:Slovak}{
     Astronomical Institute of the Slovak Academy of Sciences, 05960,
     Tatranska Lomnica, the Slovak Republic}

\authorcount{affil:Sternberg2}{
  Institute of Astronomy, Russian Academy of Sciences,
  119017 Pyatnitskaya Str., 48, Moscow, Russia}

\authorcount{affil:Miller}{
     Furzehill House, Ilston, Swansea, SA2 7LE, UK}

\authorcount{affil:Neustroev}{
     Astronomy Research Unit, PO Box 3000,
     FIN-90014 University of Oulu, Finland}

\authorcount{affil:Neustroev2}{
     Finnish Centre for Astronomy with ESO (FINCA), University of Turku, 
     V\"{a}is\"{a}l\"ntie 20, FIN-21500 Piikki\"{o}, Finland}

\authorcount{affil:Neustroev3}{
Instituto Nacional de Astrof{\'i}sica, {\'O}ptica y Electr{\'o}nica,
Luis Enrique Erro 1, Tonantzintla, Puebla, 72840, M{\'e}xico
}

\authorcount{affil:George1}{
  American Assosication of Variable Star Observers, 49 Bay State Road, Cambridge, MA 02138, USA}

\authorcount{affil:George2}{
  The George-Elma Observatory, Mayhill, New Mexico, USA}

\authorcount{affil:Mdy}{
     Kaminishiyamamachi 12-14, Nagasaki, Nagasaki 850-0006}

\authorcount{affil:Ioh}{
     VSOLJ, 1001-105 Nishiterakata, Hachioji, Tokyo 192-0153}

\authorcount{affil:Masi}{
     The Virtual Telescope Project, Via Madonna del Loco 47, 03023
     Ceccano (FR), Italy}

\authorcount{affil:UNAM}{
     Instituto de Astronom\'{\i}a UNAM, Apartado Postal 877, 22800 Ensenada
     B.C., M\'{e}xico}

\authorcount{affil:Dubovsky}{
     Vihorlat Observatory, Mierova 4, Humenne, Slovakia}

\authorcount{affil:Kiyota}{Variable Star Observers League in Japan (VSOLJ), 7-1 Kitahatsutomi, Kamagaya, 
Chiba 273-0126}

\authorcount{affil:trd}{Polaris Observatory, Hungarian Astronomical Association, Laborc u. 2/c, 1037 Budapest, Hungary}

\authorcount{affil:Nyrola}{
     Hankasalmi observatory, Jyvaskylan Sirius ry, Verkkoniementie
     30, 40950 Muurame, Finland}

\authorcount{affil:Ruiz1}{
     Observatorio de C\'antabria, Ctra. de Rocamundo s/n, Valderredible, 
     39220 Cantabria, Spain}

\authorcount{affil:Ruiz2}{
     Instituto de F\'{\i}sica de Cantabria (CSIC-UC), Avenida Los Castros s/n, 
     E-39005 Santander, Cantabria, Spain}

\authorcount{affil:Ruiz3}{
     Agrupaci\'on Astron\'omica C\'antabria, Apartado 573,
     39080, Santander, Spain}

\maketitle
\begin{abstract}
We report on a superoutburst of a WZ Sge-type dwarf nova (DN), ASASSN-15po. 
The light curve showed the main superoutburst and multiple rebrightenings.
In this outburst, we observed early superhumps and growing (stage A) superhumps 
with periods of 0.050454(2) and  0.051809(13) d, respectively.
We estimated that the mass ratio of secondary to primary ($q$) is 0.0699(8)
by using $P_{\rm orb}$ and a superhump period $P_{\rm SH}$ of stage A.
ASASSN-15po [$P_{\rm orb} \sim$ 72.6 min] is the first DN with the orbital period between 67--76 min.
Although the theoretical predicted period minimum $P_{\rm min}$ of hydrogen-rich cataclysmic variables (CVs)  is about 65--70 min,
the observational cut-off of the orbital period distribution at 80 min implies that 
the period minimum is about 82 min, and the value is widely accepted.
We suggest the following four possibilities: the object is (1) a theoretical period minimum object (2) a binary with a evolved secondary (3) a binary with a metal-poor (Popullation II) seconday (4) a binary which was born with a brown-dwarf donor below the period minimum.
\end{abstract}

\section{Introduction}\label{sec:int}
Cataclysmic variables (CVs) are close binary systems which are composed of 
a white dwarf (WD) primary and a Roche-lobe-filling secondary.
The transfered mass from the secondary forms an accretion disk around the primary.
DNe are a subclass of CVs characterized by a sudden brightening of the disk, called outburst.
Outbursts are thought to be caused by thermal instability of disk (see e.g. \cite{war95book}).

\begin{figure}
\begin{center}
\FigureFile(80mm,50mm){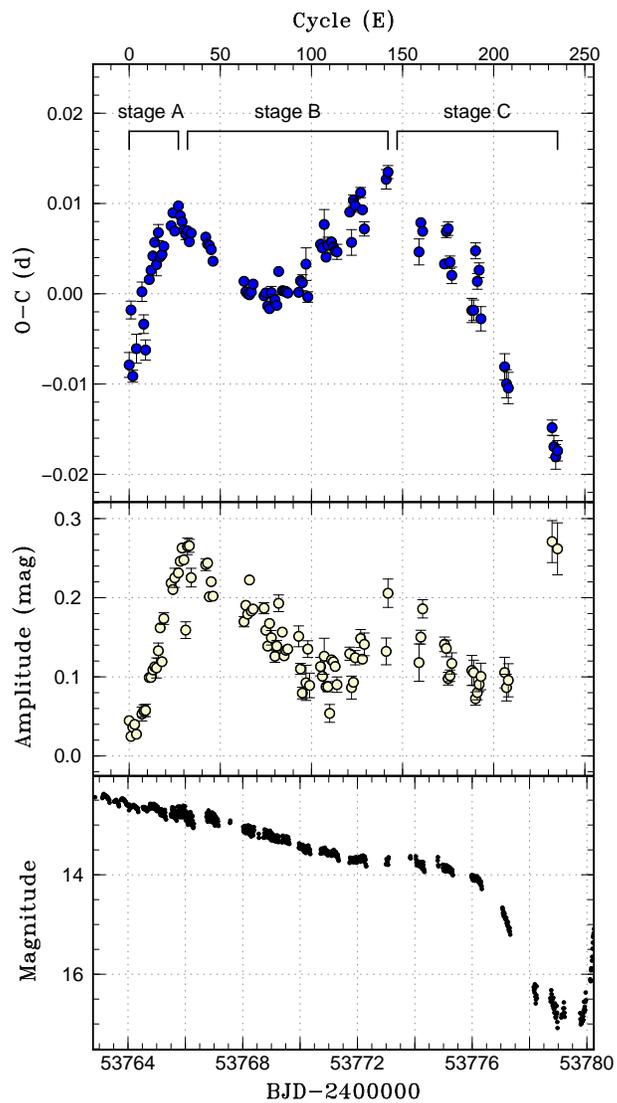}
\end{center}
\caption{Examples of the classification of superhumps. Figure is taken from \citet{mkimura}, and the plotted data is the outburst of ASAS J102522-1542.4 during 2006 which was analyzed in \citet{kato2012temp}. Upper: $O-C$ diagram. Middle: superhump amplitude. Lower: overall light curve. The classification is defined by the variations of the period and amplitude.}
\label{fig:shtemplate}
\end{figure}

SU UMa-type dwarf novae (DNe) show not only normal outbursts but also superoutbursts which
are caused by thermal-tidal instability \citep{osa89suuma,osa96review}. 
In superoutbursts, superhumps can be observed which have a small amplitude of 0.1--0.5 mag with
a period a few percent longer than $P_{\rm orb}$. 
Superhumps are considered to be a result of the 3:1 resonance of accretion disks 
which makes disks elliptical \citep{whi88tidal,lub91SHa,lub91SHb,hir90SHexcess}.
According to \citet{Pdot}, ordinary superhumps are classified to three stages by how the period changes: 
stage A, stage B, and stage C (see Figure \ref{fig:shtemplate} on the classification of superhumps). 
The stage A superhumps are thought to represent the growing phase of the 3:1 resonance, 
and their period is considered to reflect the disk precession rate at the 3:1 resonance radius.
The period of stage B superhumps is shorter because of pressure effects in the disk,
and the origin of stage C superhumps is unclear (for more detail, see \cite{kat13qfromstageA}).

WZ Sge-type DNe are a subclass of SU UMa-type DNe which shows few normal outbursts and, 
compared with SU UMa-type DNe, have especially rare outbursts and short orbital periods. 
WZ Sge-type DNe feature double-peak variations in their light curves, called early superhumps, which are observed in the initial term of superoutbursts. 
It is believed that the period of early superhumps corresponds reasonably well with the orbital period \citep{kat02wzsgeESH,ish02wzsgeletter}. 
They also exhibit multiple rebrightenings after the main superoutburst (see \cite{kat15wzsge} for review).
The reason why early superhumps are observed in WZ Sge-type DNe is that 
the binaries have extremely small mass ratios $q$ , so the disks can spread to the 2:1 resonance radius \citep{osa02wzsgehump}. 
Once the 2:1 resonance is excited, two-armed dissipation patterns appear near the 2:1 resonance radius, 
and we see the geometrical superposition of two light sources.
This is how early superhumps are observed with double-peaked profiles (cf. \cite{mae07bcuma}). 
The 2:1 resonance is thought to suppress the growth of the 3:1 resonance \citep{lub91SHa}. Hence, after the 2:1 resonance becomes weak, the 3:1 resonance becomes excited and ordinary superhumps grow.
There are several cases when rebrightenings occur after main superoutbursts. 
The rebrightenings are grouped into five types: only one long duration rebrightening (type A), 
multiple rebrightenings (type B), only one short duration rebrightening (type C), no rebrightenings (type D), 
and double superoutburst (type E) \citep{ima06tss0222,Pdot,kat13j1222}.
The rebrightening types appears to reflect the evolutionary phase, and
the order of the evolution seems to be type C$\to$D$\to$A$\to$B$\to$E as \citet{kat15wzsge} suggested.

According to the evolutionary theory of CVs, a binary separation become shorter mainly
by magnetic braking in initial stage (orbital period $P_{\rm} >$ 3 hrs), 
and by gravitational wave radiation (GWR) in the last stage (orbital period $P_{\rm orb} <$ 3 hrs) \citep{pac81CVevolution}. 
As mass transfer from the secondary proceeds, the secondary becomes partially degenerate. 
The smaller the mass of the degenerate secondary star becomes, the larger its radius becomes. 
Such a transition of the mass-radius relation causes the increasing orbital period. 
Hence it is believed that there is a theoretical lower limit of the period of CVs, called the period minimum $P_{\rm min}$
(\cite{pac81CVGWR,kin88binaryevolution}).
Assuming that the angular momentum loss of the binary is driven purely by the GWR,
the theoretical $P_{\rm min}$ is calculated to be 65--70 min \citep{kol99CVperiodminimum,how01periodgap}. 
However, the observational cut-off of the $P_{\rm orb}$ distribution is about 80 min,
and the peak, called the period spike, is about 82 min \citep{gan09SDSSCVs}.
Thus, \citet{kni11CVdonor} suggested an increased rate of the angular momentum loss 
to match the observational distribution, and reproduced $P_{\rm min}$ of 82 min.
Although the value of 82 min has been widely accepted, the mechanism of the larger angular momentum loss is unclear,
and the discrepancy between the theory and observation is not settled yet (period minimum problem).
CVs below the period minimum are generally classified as evolved objects called EI Psc-type, or helium-rich objects called AM CVn-type.

\citet{szk05SDSSCV4}, \citet{lit07OVBoo} and \citet{pat08j1507} reported on 
the peculiar ``WZ Sge-like'' CV OV Boo (SDSS 150722.30+523039.8) whose orbital period of 67 min is below the period minimum of 82 min.
``WZ Sge-like'' means that this object has shown no outbursts but 
the spectroscopy suggests a hydrogen-rich CV and
the binary parameters, measured by the eclipsing light curve, are similar to WZ Sge-type DNe.
To explain its short orbital period, 
they suggested the possibility that the secondary is a metal-poor (Population II) star.
We do not know, however, whether an object like OV Boo shows outbursts like WZ Sge-type objects. 

In this paper, we report on a new WZ Sge-type DN below the period minimum, ASASSN-15po. 
It was first detected in superoutburst at $V$=13.7 on 2015 September 20.46 UT by All-Sky Automated Survey for SuperNovae (ASAS-SN, \cite{ASASSN}),
using data from Brutus telescope in Hawaii \citep{atel8042}.
Its position is (RA:) 00h36m35.83s, (Dec:) +21$^\circ$51'25''.7. The quiescent counterpart was $g$=21.6 mag SDSS J003635.82 +215125.7, and it was reported to be fainter than $V$=17.6 on 2015 September 8.47 UT.
This object showed early superhumps and multiple rebrightenings, which suggest that the object is a WZ Sge-type DN.
Measuring $P_{\rm orb} \sim$ 72.6 min by using the period of the early superhumps,
we found out that the object has a very short orbital period below the commonly accepted period minimum of 82 min. 
We introduce observational and analysis methods in section 2, show the results of the analysis in section 3, and discuss the results in section 4.

\section{Observation and Analysis}\label{sec:ana}
ASASSN-15po was observed by several observers, and the log of photometric observations is listed in table 1.
We also used the public data from AAVSO International Database\footnote{
   $<$http://www.aavso.org/data-download$>$.
}.
Before analyzing the data, we applied zero-point corrections to each observer by adding constants.
Moreover, we converted time to barycentric Julian date (BJD).
When we analyzed the period of superhumps or early superhumps, 
we subtracted a global variation of superoutburst by using locally weighted polynomial regression fitting, LOWESSFIT \citep{LOWESS}, 
and calculated periods by phase dispersion minimization (PDM) analysis \citep{PDM}.
The 1$\sigma$ error of the PDM method was estimated by the same method of \citet{fer89error} and \citet{Pdot2}.
The $O-C$ diagrams were drawn to search the variation of the superhump periods in the same way as \citet{Pdot}.

We also obtained a low S/N optical spectrum.
On BJD 2457311.79, ASASSN-15po was observed at the Guillermo Haro Observatory at Cananea, Sonora, Mexico on the 2.1-m telescope with the Boller \& Chivens spectrograph, equipped with a 24~$\mu$m ($1024\times1024$) Tektronix TK1024 CCD chip. 
The observation was taken in the wavelength range of 3900--7150 {\AA} with a dispersion of 3.2~{\AA} pixel$^{-1}$. 
The corresponding spectral resolution was about 6.5~\AA. 
We took one spectrum with 30~min exposure time. 
The weather was clear during the observation, and the seeing was around $\sim 1.5$\arcsec. 
The reduction procedure was performed using {\sc iraf}. 
A comparison spectrum of a He-Ar lamp was acquired for the wavelength calibration.

\section{Result}\label{sec:res}
\subsection{Overall Light Curve}
\begin{figure}
\begin{center}
\FigureFile(80mm,50mm){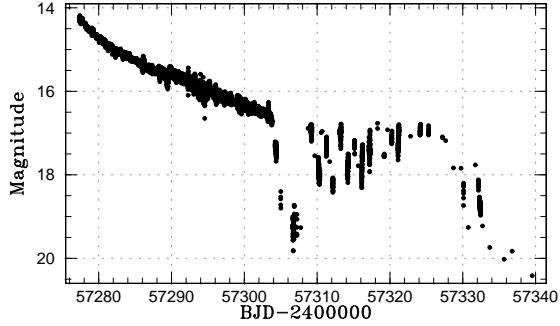}
\end{center}
\caption{Overall light variation ASASSN-15po. 
The superoutburst and multiple rebrightenings are visible. 
All data are binned to 0.01 day.}
\label{fig:lc}
\end{figure}

The superoutburst of ASASSN-15po was detected at $V$=13.7 on BJD 2457275.94 by the ASAS-SN team, 
and the superoutburst was seen as in figure \ref{fig:lc}. 
In quiescence, this object was $g=$ 21.6 in SDSS data \citep{atel8042},
so the superoutburst had a large amplitude of at least $\sim$8 mag.
Time-resolved photometry started on BJD 2457277.38, 
and early and ordinary superhumps were observed during the plateau phase.
Following the main superoutburst, which lasted for about 28 days until BJD 2457303.32, 
there were multiple rebrightenings between BJD 2457309.10 and BJD 2457332.46.
This light curve represents the type A/B (multiple) rebrightenings according to \citet{kat15wzsge}.
Type A/B is often seen in WZ Sge-type DNe evolved around or beyond the period minimum
(for more detail about the discussion of the rebrightening types of WZ Sge-type DNe, see \citet{kat15wzsge}).

\subsection{Spectroscopy}\label{sec:spectrum}
\begin{figure}
\begin{center}
\FigureFile(80mm,50mm){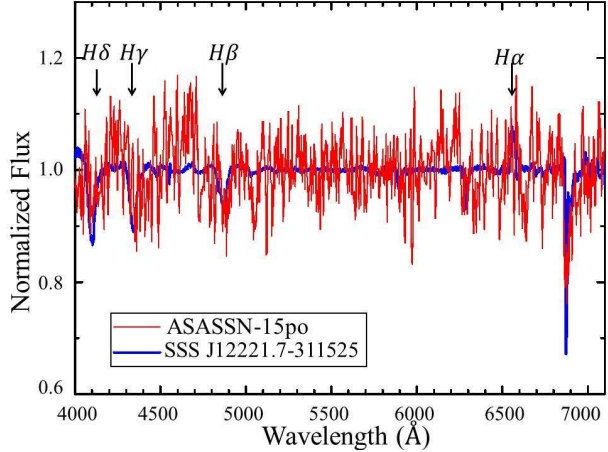}
\end{center}
\caption{
This Figure shows the normalized spectrum observed on BJD2457311.79 (during rebrightenings). 
The red spectrum is ASASSN-15po and the blue is SSS130101:122221.7-311525 (SSS122222).
It was observed at the Guillermo Haro Observatory at Cananea, Sonora, Mexico on the 2.1-m telescope with the Boller \& Chivens spectrograph, equipped with a 24~$\mu$m ($1024\times1024$) Tektronix TK1024 CCD chip. 
The observation was taken in the wavelength range of 3900--7150 {\AA}, 
and the spectral resolution was about 6.5~\AA.}  
\label{fig:spectrum2}
\end{figure}

The spectrum of ASASSN-15po was obtained during rebrightnings on BJD2457311.79, and the normalized spectrum is in Figure \ref{fig:spectrum2}. 
The red spectrum is ASASSN-15po and the blue is SSS130101:122221.7-311525 (taken from Neustroev et al. (2016)) which is known as a typical WZ-Sge type novae.
The spectrum of ASASSN-15po shows $\rm H\beta$, $\rm H\gamma$, and $\rm H\delta$ absorption lines, which are typical properties of that of WZ-Sge type novae during the later part of superoutbursts.

\subsection{Early Superhumps}\label{sec:earlySH}
\begin{figure}
\begin{center}
\FigureFile(80mm,50mm){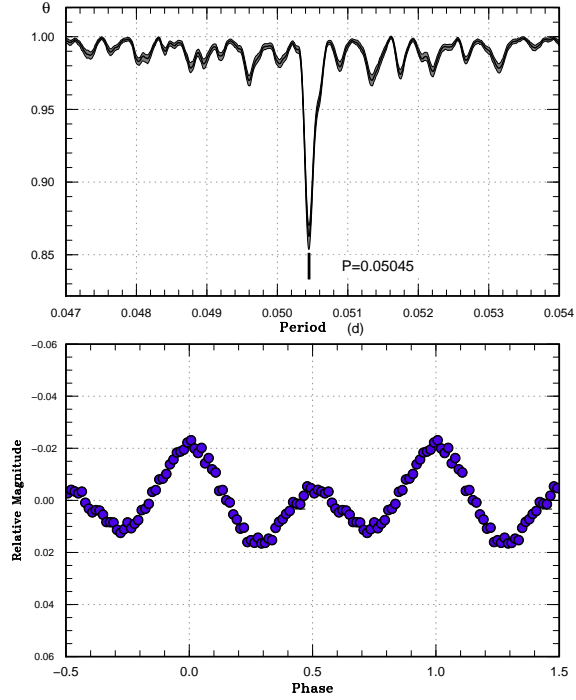}
\end{center}
\caption{The upper figure is the result of the PDM analysis of early superhumps, and the period is 0.050454(2) d. 
The lower figure is phase averaged profile of early superhumps. The profile clearly shows the double-wave modulations.}
\label{fig:pdme}
\end{figure}

\begin{figure}
\begin{center}
\FigureFile(80mm,50mm){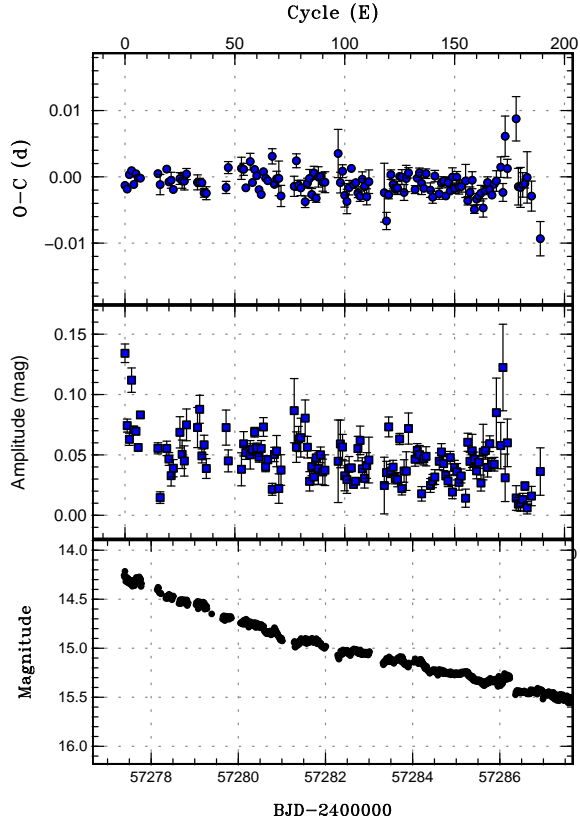}
\end{center}
\caption{$O-C$ analysis of early superhumps. The upper figure is an $O-C$ diagram of early superhumps (BJD 2457277.38 -- 2457285.62), 
and one can see that the period is almost constant. 
The middle figure shows the variation of the amplitude. The lower figure is the light curve.}
\label{fig:oce}
\end{figure}

\begin{figure}
\begin{center}
\FigureFile(80mm,50mm){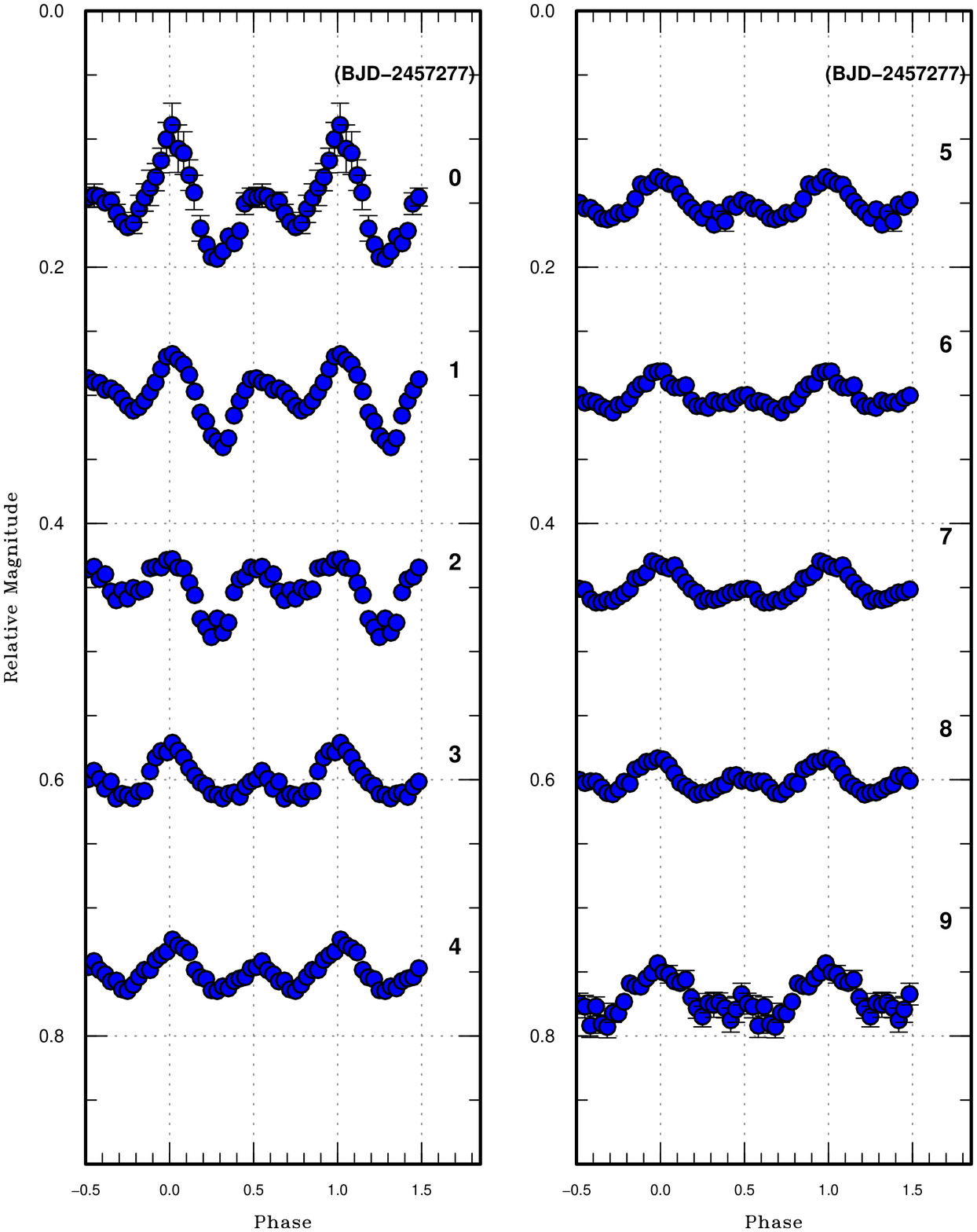}
\end{center}
\caption{Daily variation of the phase-averaged profiles of early superhumps. The numbers at the right end of each light curve represent the number of days elapsed since BJD 2457277.}
\label{fig:time}
\end{figure}

As we mentioned in section \ref{sec:int}, it is believed that the 2:1 resonance is 
the cause of early superhumps \citep{osa02wzsgehump}, and it is usually observed as a double-peaked modulation. 
It is well-known that the orbital period is very close to the period of early superhumps to an accuracy of 0.1$\rm \%$ \citep{kat15wzsge}.
We observed the early superhumps in this superoutburst (BJD 2457277.38--2457285.62), 
and we regarded the period as the orbital period $P_{\rm orb}$. 
The result of PDM analysis of the early superhumps is 0.050454(2) d (in figure \ref{fig:pdme}). 
As one can see in figure \ref{fig:time}, the phase-averaged profiles changed day by day.
As in figure \ref{fig:oce}, the $O-C$ diagram during early superhumps indicated that the period is almost constant.
The times of the early superhump maxima, which were used to draw the $O-C$ diagram, are listed in table 2.


\subsection{Ordinary Superhumps}
Ordinary superhumps can be seen as a result of the 3:1 resonance.
We drew the $O-C$ diagram in the same way in section \ref{sec:earlySH}.
The times of the superhump maxima, which were used to draw the $O-C$ diagram, are listed in table 3.
In figure \ref{fig:ocab}, we can see the clear phase transition at $E=$ 48.
We determined that stage A occurred from BJD 2457286.97--2457289.46 ($E \leq$ 48) 
because of the increasing amplitude (see the middle panel of figure \ref{fig:ocab}),
and stage B lasted between BJD 2457289.46--2457303.79 ($E =$ 49--329) because of the gradual variation of $P_{\rm SH}$ and the decreasing amplitude.
There seems to be no stage C because the main superoutburst finished before the appearance of stage C.
In stage A and B, we obtained the periods of 0.051809(13) and 0.050913(2) d, respectively, by the PDM analysis. 
The rate of variation of the period in stage B ($P_{\rm dot} \equiv \dot{P_{\rm SH}} / P_{\rm SH}$) is $\rm 1.29(17)\times10^{-5}$.
After the main superoutburst, we could not see clear superhumps 
because of the sparse data and the interference by rapid variations of rebrightenings.

\begin{figure}
\begin{center}
\FigureFile(80mm,50mm){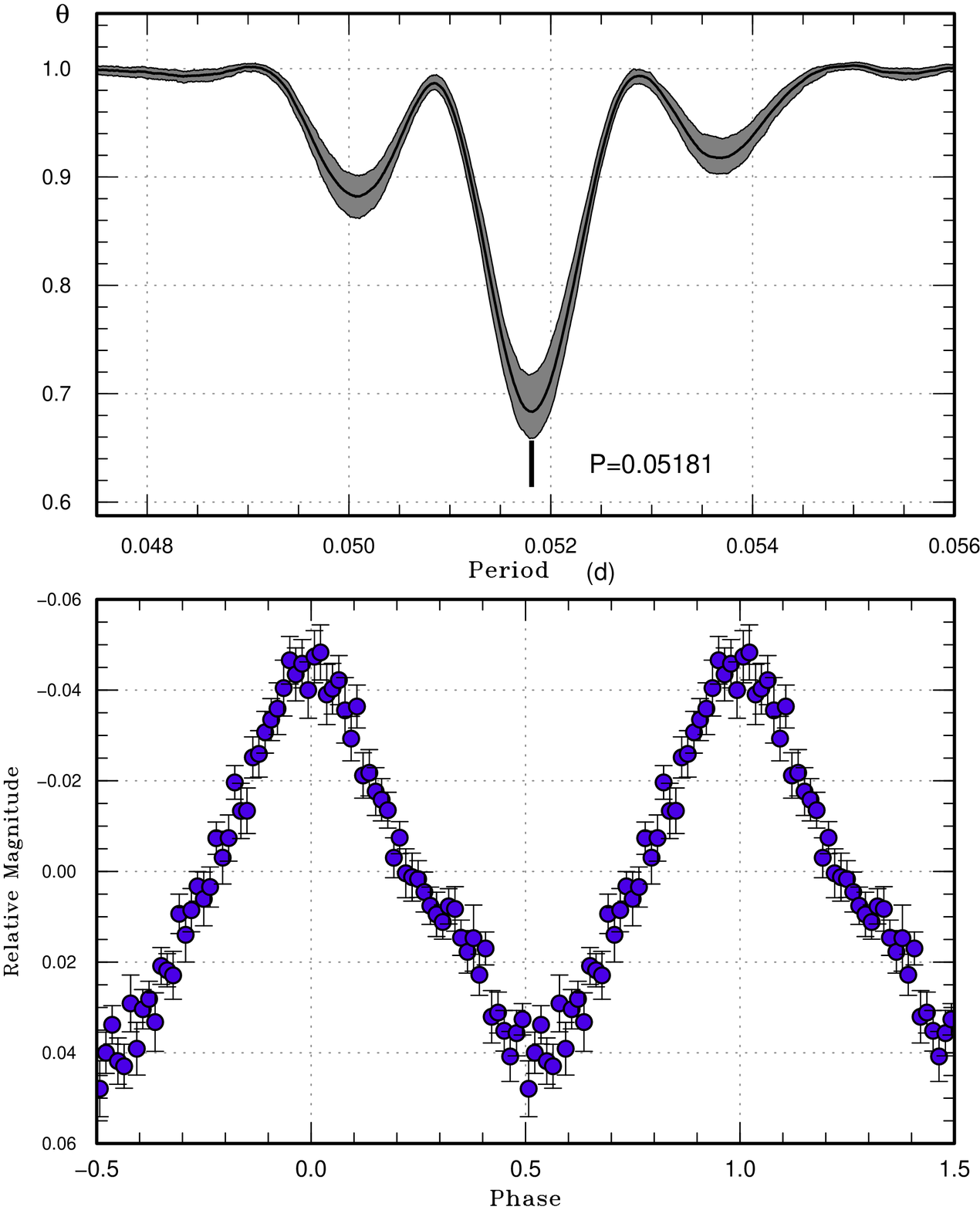}
\end{center}
\caption{The upper figure is the result of PDM analysis of stage A superhumps (BJD 2457286.55 -- 2457288.63), and the period is 0.051809(13). 
The lower figure is phase averaged profile.}
\label{fig:pdma}
\end{figure}
\begin{figure}
\begin{center}
\FigureFile(80mm,50mm){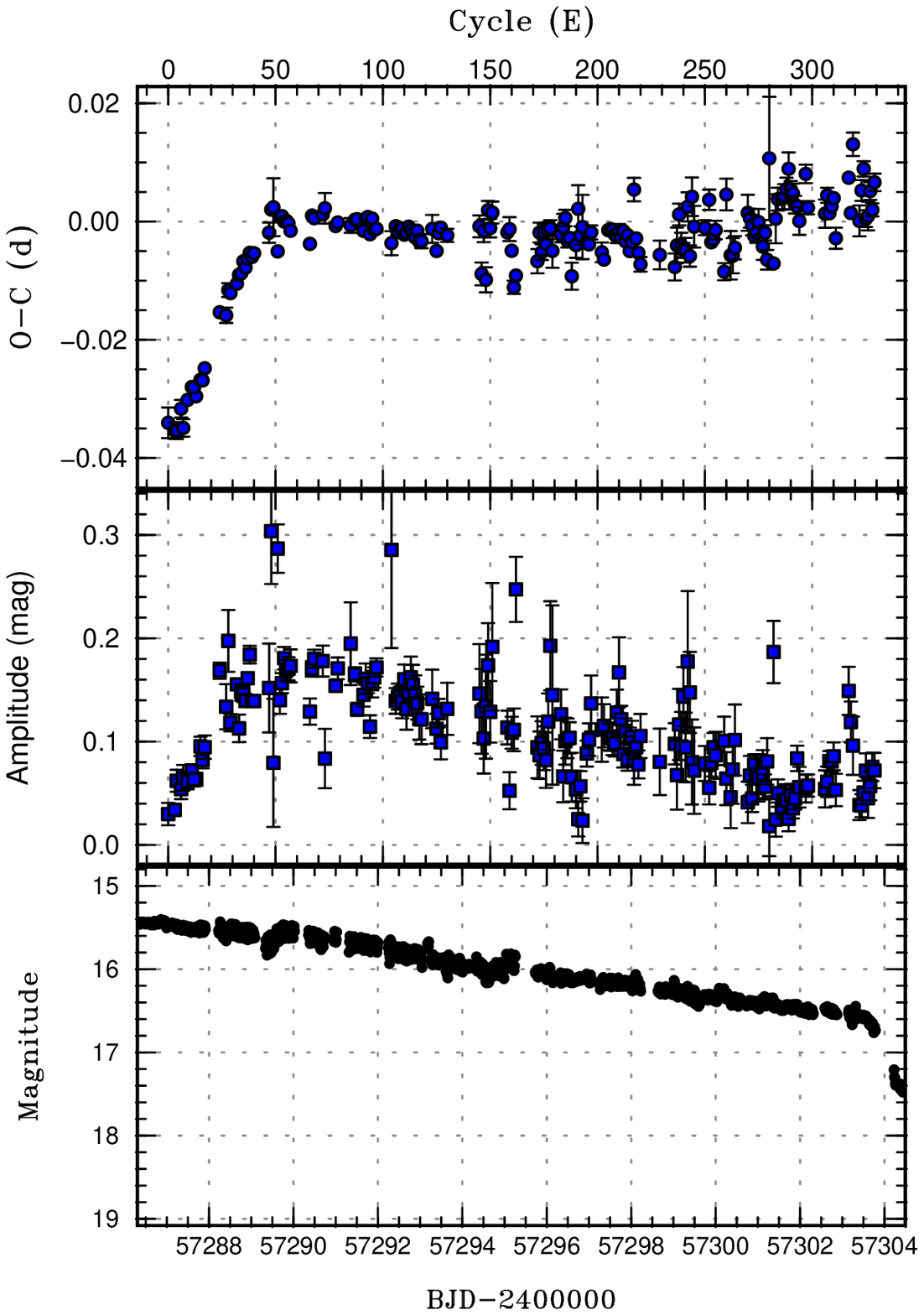}
\end{center}
\caption{The upper figure is the $O-C$ diagram of ordinary superhumps (BJD 2457286.55 -- 2457303.32), 
and one can clearly see the transition between stage A and B at $E=$ 48. 
The middle figure shows the variation of amplitudes, and the lower figure is magnitude.}
\label{fig:ocab}
\end{figure}

\section{Discussion}\label{sec:dic}
\subsection{Orbital Period below the Period Minimum}
Orbital periods of CVs are very important for determining the evolutionary stage of CVs. 
As mentioned in the previous section, we regarded the period of early superhumps as the orbital period, 
and the value is 0.050454(2) d ($=$ 72.6 min) below the period minimum of 82 min.
ASASSN-15po is the first DN with an orbital period between 67--76 min
and the first WZ Sge-type DN below the commonly accepted period minimum
except for the ultra-short orbital period objects, 
which are helium-rich CVs like AM CVn-type or EI Psc-type CVs or their candidates \citep{bre12j1122}.
Although this orbital period is unusual for an ordinary CV,
the profile of the superoutburst is very similar to ones of ordianry WZ Sge-type DNe.

\subsection{Mass Ratio}
Mass ratios of secondary star to primary star ($q=M_{\rm 2}/M_{\rm 1}$) are one of the most important properties for discussing the evolution of binaries. 
It had been estimated as a function of fractional superfump excess ($\epsilon= \omega_{\rm pr}/ \omega_{\rm orb}-1$) using an empirical relation derived by \citet{pat05SH}. This method, however, suffers from the degree of pressure effect \citep{pearsonSHN}. Recently, deepened understanding of precession disk enable us to develop the estimation method of mass ratio of SU UMa-type DNe.
By using the method of \citet{kat13qfromstageA},
we can estimate the mass ratio $q$ from $P_{\rm orb}$ and $P_{\rm SH}$ of stage A.
We define the fractional superhump excess $\epsilon^{\ast}$ as 
\begin{equation}
\epsilon^{\ast}= \omega_{\rm pr}/ \omega_{\rm orb} = 1-P_{\rm orb}/P_{\rm SH},
\end{equation}
where $\omega_{\rm pr}$ is the aspidal precession rate of the eccentric disk,
and $\omega_{\rm orb}$ is the orbital angular frequency.The dynamical precession rate $\omega_{\rm dyn}$ at the radius $r$ is as follows (Note that equation \ref{eq:omega} in \citet{kat13qfromstageA} is a misprint \citep{katoarxiv}.)
\begin{equation}\label{eq:omega}
\frac{ \omega_{\rm dyn} }{ \omega_{\rm orb} } = \frac{q}{\sqrt{1+q}} \left( \frac{\sqrt{r}}{4} b_{3/2}^{(1)} \right),
\end{equation}
where $r$ is normalized by the binary separation and $\frac{1}{2}b_{s/2}^{(1)}$ is the Laplace coefficient \citep{hir90SHexcess},
\begin{equation}
\frac{1}{2} b_{s/2}^{(j)}(r) = \frac{1}{2\pi} \int_0^{2 \pi} \frac{ \cos{(j\phi)}d\phi }{ (1+r^2-2r \cos{\phi})^{s/2} }
\end{equation}
We know that the dynamical precession during stage A occurs at the 3:1 resonance radius,
which is given by
\begin{equation}
r_{\rm 3:1} = 3^{(-2/3)} (1+q)^{-1/3}.
\end{equation}
By numerically solving the above equations, the mass ratio ($q$) can be obtained just from $P_{\rm orb}$ and $P_{\rm SH}$ of stage A.
We calculated that the $q$ of ASASSN-15po is 0.0699(8).
This is a normal value in WZ Sge-type DNe.

We also know the empirical relation of WZ Sge-type DNe 
between $q$ and $P_{\rm dot} \equiv \dot{P_{\rm SH}} / P_{\rm SH}$ during stage B.
\citet{kat15wzsge} derived the following equation
\begin{equation}\label{eq:q_emp}
q = 0.0043(9) P_{\rm dot} \times 10^5 + 0.060(5).
\end{equation}
$q$ from equation \ref{eq:q_emp} is 0.066(7) which is in good agreement with $q$ from stage A.
We note that this relation is known to hold for ordinary WZ Sge-type DNe (and a part of SU UMa-type DNe),
and it is unclear whether we can use the relation for unusual WZ Sge-type DNe.
We should confirm it by measuring more unusual objects.

\subsection{Evolutionary State}
\begin{figure}
\begin{center}
\FigureFile(80mm,50mm){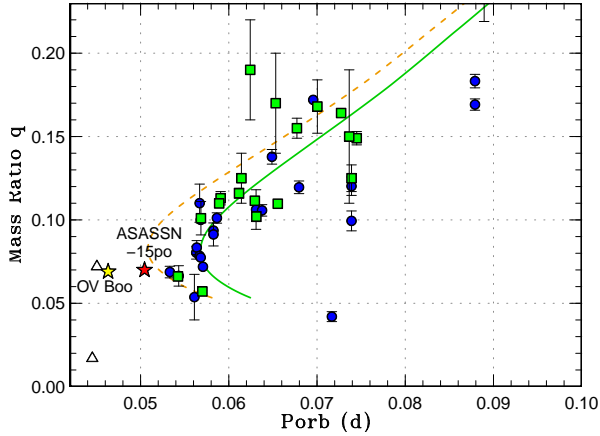}
\end{center}
\caption{The figure shows orbital period versus mass ratio of CVs, which is mainly taken from \citet{kat13qfromstageA}. Filled circles and squares are CVs for which the mass ratio are measured by stage A superhumps and eclipses, respectively. Unfilled triangles are EI Psc-type binaries, which are taken from Oshima et al. (in prep).  The dashed line is one of the standard evolutionary track, and the thick line is the modified one. ASASSN-15po and OV Boo are also plotted with star symbols.}
\label{fig:evo}
\end{figure}

We introduced the evolutionary theory of CVs in section 1, 
and the evolutionary track can be clearly seen 
on the relation between mass ratio $q$ versus $P_{\rm orb}$ in figure \ref{fig:evo}, 
which is taken from \citet{kat13qfromstageA}.
Filled circle and squares are CVs for which mass ratios have been are measured 
by stage A superhumps and eclipses, respectively.
The dashed line is one of the standard evolutionary tracks, 
which is assumed that the angular momentum loss is driven only by the GWR, 
and the thick line is another one, which is modified by assuming a higher angular momentum loss to correspond with the observational studies,
with the period minimum of 82 min from \citet{kni11CVdonor}.
The mechanism of the higher angular momentum loss rate is unclear;  
nevertheless, the modified evolutionary theory has been widely accepted.
The scattering around $P_{\rm orb}\sim$0.074 is mainly caused by observational difficulties. One system (IY UMa) with $q\sim$0.10 is an eclipsing one and it is difficult to determine the period of stage A superhumps due to overlap with eclipses (see, \cite{kat13qfromstageA}). Other systems with relatively long $P_{\rm orb}$ ($\sim$0.074) also have difficulties since stage A superhumps in high-$q$ systems last only 1-2d, in contrast to several days in low-$q$ systems. 
Although the majority of CVs are considered to evolve along the evolutionary track as the mass transfer proceeds,
ASASSN-15po is located below the period minimum. 
This object is a long-sought candidate of the theoretical period minimum CV, which evolves purely by the GWR.

This possibility, however, may be less likely since the majority of known CVs are not apparently on the theoretical evolutionary track and there is no physical reason why one single CV out of more than 2000 should undergo a different type of angular momentum loss.
The systems with high-density secondaries can have shorter orbital periods below the period minimum. 
As we consider the Roche-Lobe geometry, the relation of $P_{\rm orb}$ and 
density of a secondary star $\rho_2$ is restricted to $P_{\rm orb} \sqrt{\rho_2}=constant$ \citep{fau72amcvn}. 
From this relation, we can say that the shorter orbital period the system has, the higher density the secondaries should be. 
Hence, the short orbital period of ASASSN-15po suggests that this system has a compact, high-density secondary.

One explanation to the high-density secondary is that the system has a slightly evolved secondary, 
which are called EI Psc-type CVs \citep{fau72amcvn}.
The density of evolved stars is higher than that of hydrogen-rich stars \citep{fau72amcvn}.
Recently, several EI Psc-type CVs have been discovered and are thought to be intermediate objects evolving toward the AM CVn-type systems in the CV channel.
The identification of EI Psc-type is an unusually hot donor for the short orbital period \citep{tho02j2329}, which can be explained by an evolved donor that has been stripped of its hydrogen layers. Another hallmark of these system is that they are N -enhanced, and C-depleted \citep{eipscpap}.
There is, however, only a low-resolution spectroscopic observation of ASASSN-15po (see section \ref{sec:spectrum}), 
so we cannot determine whether ASASSN-15po is classified as a EI Psc-type CV.
It is known that evolved CVs often show profiles of superoutbursts different from ones of hydrogen-rich CVs 
because of the difference of the ionization temperature between the hydrogen and evolved helium-rich disks \citep{tsu97amcvn}.
However, it is known that some EI Psc-type CVs show outbursts similar to typical hydrogen-rich objects,
because the secondaries of such objects are considered to be only slightly evolved 
and have a hydrogen-rich surface and a helium-rich core.
The disk composition of such objects is, thus, still sufficiently hydrogen-rich (Ohshima et al. in preparation).
Considering this, there is a possibility that ASASSN-15po is a member of the EI Psc-type CVs.
We are not sure, however, whether the surface and core composition is sufficiently different since these stars may be fully, or nearly fully convective.

The other possibility is that this object is metal-poor (Population II). 
As discussed in \citet{pat08j1507}, if the metallicity of the secondary is poor, the opacity of the surface is also small, 
which leads to about a 20$\%$ smaller radius of the secondary.
Lower metallicity enables the period minimum to become shorter \citep{ant87KukarkinParenago}. 
Hence, the short orbital period of ASASSN-15po can imply that the object belongs to Population II. 
However, we have to note that, as mentioned in section \ref{sec:int}, CVs with low metallicity may not undergo the 
same outbursts as CVs with normal metallicity because of the different opacity of the accretion disk.
In ordinary hydrogen-rich CVs, the middle branch of the S-curve mainly represents
the opacity variation due to the partial ionization of hydrogen. It is known that iron-group
elements contribute to this opacity, and we may expect some difference in the S-curve
between Population I CVs and iron-poor Population II ones, and consequently some difference in the outburst behavior (cf. different S-curve in different metallicities in \citet{pojm1986}). This metallicity effect becomes more prominent in hydrogen-poor systems (fig.3 in
\cite{tsu97amcvn}). We do not have observational evidence whether outburst properties
are different in Population II CVs from Population I ones, but since the number of CVs in
globular clusters is increasing (cf. \cite{bell2016}), we may have
observational evidence in near future how Population II CVs behave during outbursts.
Here, we showed in section \ref{sec:res} that ASSASN-15po behaved very similarly to ordinary hydrogen-rich CVs. 
Assuming the difference of the outburst behavior between Population I and II,
our observations suggest that ASASSN-15po is not a Population II star.
We have no way to identify whether the secondary belongs to Population II because there is no data of proper motion of this object.
In order to check this interpretation, we need a theoretical model calculation of DN outbursts for different metallicities and a direct spectroscopic observation in quiescence.

Finally, we have to consider a possibility that the binary was born with a brown-dwarf donor below the period minimum. This system may have started mass transfer below period minimum, and evolved toward present longer period.

\section{Summary}
We reported photometric observations and a low-resolution spectroscopy of the superoutburst of ASASSN-15po.
The light curve showed a typical WZ Sge-type superoutburst with the multiple rebrightenings, classified as type A/B.
The main superoutburst lasted for about 28 days, and, after a few days, 
the rebrightening phase began and continued for at least 24 days.
We could also see early superhumps (BJD 2457277.38--2457285.62), 
stage A superhumps (BJD 2457286.97--2457289.46) and stage B superhumps (2457289.46--2457303.79).
Superhump periods are 0.050454(2), 0.051809(13) and 0.050913(2) d, respectively.
The rate of variation of the period in stage B ($P_{\rm dot}$) was calculated to be $\rm 1.24(18)\times10^{-5}$.
By using $P_{\rm orb}$ and $P_{\rm SH}$ of the stage A, we estimated $q=$ 0.0701(8).
We also estimated $q=$0.066(7) by using an empirical relation between $q$ and $P_{\rm dot}$ for normal WZ Sge-types.
The latter $q$ is in good agreement with the former $q$.

We found that ASASSN-15po has an orbital period below the observational period minimum of 82 min.
Although the evolutionary theory suggests that the period minimum is 65--70 min,
there are no DNe with the orbital period between 67--76 min.
We suggested four possibilities:
\begin{itemize}
\item This object is a theoretical period minimum object whose angular momentum is extracted purely by the GWR.
\item The secondary of ASASSN-15po has been partially stripped, with an evolved, helium-rich core and hydrogen-rich surface atmosphere.
\item The secondary is a metal-poor object belonging to Population II.
The low opacity of a Population II star leads to a smaller radius and higher density --- and thus a shorter period minimum.
\item The binary was born with a brown-dwarf donor below the period minimum.
\end{itemize}
To determine the status of ASASSN-15po, we have to carry out a detailed spectroscopic observation.

\bigskip
Acknowledgements:
We are grateful to many worldwide observers 
who have contributed to AAVSO International Database for observing and providing data of cataclysmic variables. 
We acknowledge with thanks ASAS-SN team for surveying and making the information available to the public.
We also acknowledge very helpful comments from the referee.
This work was supported by a Grant-in-Aid ``Initiative for High-
Dimensional Data-Driven Science through Deepening of Sparse
Modeling'' from the Ministry of Education, Culture, Sports,
Science and Technology (MEXT) of Japan (25120007). 
It was also partially supported by grants RFBR No 15-02-06178, RFBR No 14-02-0082 (S.S.), VEGA No. 2/0002/13 (S.S.) and RSF No 14-12-00146 (P.G., for processing observation data from Slovak observatory).


\newcommand{\noop}[1]{}

\end{document}